\documentclass[twocolumn,superscriptaddress,showpacs,preprintnumbers,amsmath,amssymb]{revtex4}
\usepackage{graphicx}
\usepackage{textcomp}   
\usepackage[scaled]{helvet}  
\usepackage{amsmath,amssymb}  
\usepackage{type1cm}
\usepackage{bm}  
\usepackage{color}

\begin{document}

\title{Sign Flip in the Casimir Force for Interacting Fermion Systems}

\author{Antonino Flachi} 
\affiliation{Department of Physics,  Keio University, 4-1-1 Hiyoshi, Kanagawa 223-8521, Japan}
\affiliation{Research and Education Center for Natural Sciences, Keio University, 4-1-1 Hiyoshi, Kanagawa 223-8521, Japan}

\author{Muneto Nitta} 
\affiliation{Department of Physics,  Keio University, 4-1-1 Hiyoshi, Kanagawa 223-8521, Japan}
\affiliation{Research and Education Center for Natural Sciences, Keio University, 4-1-1 Hiyoshi, Kanagawa 223-8521, Japan}

\author{Satoshi Takada} 
\affiliation{Earthquake Research Institute, The University of Tokyo, 1-1-1 Yayoi, Bunkyo-ku, Tokyo 113-0032, Japan and Department of Physics, Kyoto University,\\ Kitashirakawa Oiwakecho, Sakyo-ku, Kyoto 606-8502, Japan}

\author{Ryosuke Yoshii} 
\affiliation{Research and Education Center for Natural Sciences, Keio University, 4-1-1 Hiyoshi, Kanagawa 223-8521, Japan}

\date{\today}
\begin{abstract}
In this work we consider a fermionic chain of finite length $\ell$. Fermions are allowed to interact and are forced to obey  boundary conditions, thus altering the process of condensation. Our goal is to explore how this affects the quantum vacuum energy for this system. We approach this problem by using a self-consistent method and observe a nontrivial behavior in the Casimir force, displaying a switch from an attractive to a repulsive regime. This flip stems from the competition between the attractive contribution from the usual fermionic Casimir effect and a repulsive one coming from the condensate. 
\end{abstract}

\maketitle
\section{Introduction} 
Casimir was one of the first to find a way to peek into the quantum vacuum \cite{Casimir1}. He did so by (ideally) placing two flat perfectly conducting boundaries at a separation $\ell$ in empty space. He understood that these boundaries would deform the quantum vacuum of the electromagnetic field and, as a result, experience a force. This force turned out to be attractive, scaling as $\ell^{-4}$, and macroscopically measurable \cite{Sparnay}. A few years later, Casimir used this idea for his mousetrap model of the electron that aimed at explaining the stability of charged particles from the balancing between the attractive quantum vacuum force and  the self-energy repulsion between charges \cite{Casimir2}. The model however had to be rejected, once Boyer proved that the Casimir force for a spherical shell is repulsive \cite{Boyer}. The idea had a time of rejuvenation during the 1970s with the MIT bag model of hadrons, where their stability was, this time, sought in the compensation between fermion and gauge contributions to the quantum vacuum energy \cite{MIT_bag_model}. The difference in statistics, however, did not produce any cancellation  as Johnson showed \cite{Johnson}, illustrating, once again, that the sign of the Casimir force is related to the structure of the boundaries and to the underlying field theory in a way that is difficult to anticipate.

Since Casimir's pioneering work, understanding the properties of the quantum vacuum has become a central problem in many areas of science. Examples of applications of the Casimir effect can be found in string theory \cite{Fabinger:2000jd}, brane models \cite{FlachiCasimir}, gravity \cite{Quach}, beyond-standard-model physics \cite{Decca}, nanotechnology \cite{Capasso1}, colloids \cite{dietrich}, and even in biology \cite{biocas}. The literature on the Casimir effect is vast and the reader may consult Ref.~\cite{Experiments} for some experimental results and Refs.~\cite{textbook,Milton,Klimchitskaya,Kardar:1997cu,Albrecht} for reviews, background work, and a comprehensive list of references.

In many instances, the questions revolving around the Casimir effect require an understanding on how to control the sign of the force, and eventually balance to zero the quantum vacuum energy. This was the case in Casimir's original mousetrap model of the electron, and in the MIT bag model of hadrons. In cosmology the same issue is related to the renowned cosmological constant problem \cite{Weinberg:1988cp}. In nanotechnology the very same question is reflected in the phenomena of stiction at the nanoscale that poses practical limitations to reducing the size of electromechanical systems \cite{Genet,Capasso}. 
 
A general understanding is far from complete. Some analyses showed that the force is necessarily attractive for two identical, disjoint, charge conjugate boundaries \cite{KennethBachas}, but a generic connection between the sign of the force and the properties of the system is not implied by this result. An interesting analysis was carried out by Schaden, who used a worldline approach (see Ref.\ \cite{Gies}) to inspect the dependence of the Casimir force on the shape of the boundary and considered a massless free scalar confined in a flasklike container satisfying Dirichlet boundary conditions \cite{Schaden}. In this case, competing contributions to the force originate from different types of Brownian bridges. Other ways to control and invert the sign of the Casimir force in micromechanical systems were proposed in Ref.~\cite{Munday} by introducing a suitable liquid between the boundaries. This would effectively change the boundary conditions that could be adjusted to produce a change of sign in the force. 

All the above results are nongeneric and require an externally driven tuning of the boundary conditions to induce the sign flip. This point is not new 
and, for instance, Ref.~\cite{Asorey} illustrates it for a free scalar field subject to a four-parameter family of boundary conditions. In fact, what would be ideal is a way to \textit{induce} the sign flip in the force by varying some parameters, without externally changing the boundary conditions. This could have interesting applications in all directions where the vacuum energy needs to be balanced to zero. Our goal here is to produce a simple field theoretical example where this (partially) happens. 

It should be obvious that in order to generate a sign flip in the force we need two competing contributions. One way to achieve this is to use a fermion field theory featuring condensation. For this reason we need to introduce interactions, an area that has only marginally been considered in the Casimir effect literature (see, however, Refs.~\cite{Flachi:2013bc,Schecter}; some examples of the fermion Casimir effect are in Ref.~\cite{CasFer}).  
In such a model we would then have two contributions to the vacuum energy: one coming from the condensate (the condensation energy) and the other from the fermions. At this stage it is not evident that the two contributions may compete with each other, but it is straightforward to see that they are not independent (as it can be understood by using the large-$N$ approximation and by observing that the fermion determinant depends on the condensate). This will become clear in the latter part of this Letter.  

The field theory we will use here is defined by the following $1+1$ dimensional version of the Nambu-Jona Lasinio model, also known as the chiral Gross-Neveu model \cite{NJL,Gross}: 
\begin{equation}
\mathcal{L}=i\bar \psi \slash\hspace{-0.5em}\partial \psi +\frac{g}{2} [(\bar \psi \psi)^2+(\bar \psi i\gamma_5 \psi)^2]. 
\label{eq:1}
\end{equation}
Possibly, this model offers the simplest way to incorporate our idea and it carries the added value of being exactly integrable. 
Also, it is well known that it has many applications both as a toy model of chiral symmetry breaking in QCD  \cite{Gross} as well as in a variety of condensed matter systems ranging from conductive polymers \cite{polyacetylene}, to superconductors \cite{supcond}. 
A review of many of its prominent developments and additional references can be found in Ref.~\cite{Thies:2006ti}. 

To summarize, the configuration we shall consider here consists of a linear fermionic chain of size $\ell$. Fermions are allowed to interact via a four-Fermi potential [we model the system by using the model Eq.\ (\ref{eq:1})] and are forced to obey boundary conditions at the edges of this interval. To have a physical picture, the reader may think of a string of (cold) atoms with two heavy impurities at the end points. The boundary conditions are fixed and the only parameter that can be varied is the coupling.

\section{Solutions} 
The starting point of our analysis are the fundamental equations stemming from the Lagrangian Eq.\ (\ref{eq:1}). In the large-$N$ limit, these take the following Bogoliubov-de Gennes form, 
\begin{align}
&\left(
\begin{array}{cc}
-i\partial_x & \Delta(x) \\
\Delta^\ast(x) & i\partial_x
\end{array}
\right)
\left(
\begin{array}{c}
u \\
v
\end{array}
\right)
=E\left(
\begin{array}{c}
u \\
v
\end{array}
\right),
\label{eq:2}
\end{align}
endowed with the self-consistent condition dictated by the gap equation
\begin{equation}
\Delta(x) =-g\sum_{E_n<0} u_n(x) v_n^\ast(x),
\label{eq:3}
\end{equation}
and by the boundary conditions. The above equations can be used for direct numerical integration. For the infinite line, exact solutions have been found in Refs.~\cite{Basar1,Basar2,Takahashi}. 
{It is well known that the large-$N$ approximation should be handled with care in lower dimensions. 
As argued in Ref.\ \cite{Witten}, the correlation function behaves at large distance as $\left|x\right|^{-1/N}$.
Differently from Ref.\ \cite{Witten}, here we are considering a system of finite size $\ell$.
Using the scaling behavior given above, it is straightforward to see that the present analysis is justified for $\log \ell \ll N$.}

In order to see the effect of imposing the boundary conditions, one may expand the solutions in Fourier modes. 
Straightforward steps show that the boundary conditions can be directly enforced on $\tilde u$ 
[we  define for convenience $\left(\tilde u, \tilde v\right)^T=\left(1-i\sigma_x\right)\left(u,v\right)^T/\sqrt{2}$, 
where $\sigma_x$ is the $x$ component of the Pauli matrices; 
the boundary conditions take the form $\left(1-i\sigma_x\right)\left(u,v\right)^T=0$ at the boundary], 
leading to $\tilde u=\sum_n a_n\sin \left({\pi n x}/{\ell}\right)$. 
On the other hand, the expansion for $\tilde v$ gives 
$\tilde v={E^{-1}}\sum_n a_n \left(\frac{\pi n}{\ell} \cos \frac{\pi n x}{\ell}+\Delta \sin \frac{\pi n x}{\ell}\right)$ 
that constrains the near-boundary behavior of the condensate $\Delta$ when boundary conditions are imposed.
Specifically, while $\tilde u$ is easily shown to be regular and vanishing at the boundary $x \sim 0$ consistently with the  
boundary conditions, forcing the same behavior on $\tilde v$ at $x \sim 0$ implies $\Delta(x)$ diverging as $\sim -1/x$.
The same argument can be trivially repeated near the other boundary $x=\ell$ leading to the same type of diverging behavior of the the condensate. 
This argument remains valid for both real and complex condensates. 
This near-boundary behavior of the condensate constrains the fermion fluctuations at the end points of the chain. 
The above argument shows that the imposed boundary conditions prevent the exact, 
regular solutions of Ref.\ \cite{Basar1} (BCS, kink, and Fulde-Ferrell-Larkin-Ovchinnikov) from being admissible. 

Here, we proceed by reformulating the problem in terms of a nonlinear Schr\"odinger equation, as shown in Ref.~\cite{Basar1}, 
and by requiring the solution for the condensate to satisfy the correct behavior dictated by the boundary conditions. 
For brevity, we will discuss only the main points; details will appear elsewhere.
Following Ref.\ \cite{Basar1}, the order parameter satisfies 
\begin{equation}
\Delta^{\prime\prime}+a(E)\Delta+i b(E)\Delta^\prime -2\Delta^3=0, 
\label{eq10}
\end{equation} 
with $a(E), b(E)$ constant.
The first class of real solutions satisfying the correct near-boundary behavior is
\begin{equation}
\Delta= \frac{m}{\mathsf{sn}(mx, \nu)}.
\label{bcstype}
\end{equation}
where ${\mathsf{sn}(mx, \nu)}$ and $\mathsf{cn}(mx,\nu)$ (to be used later) define the Jacobi $\mathsf{sn}$ and $\mathsf{cn}$ elliptic functions \cite{Abramowitz,Whittaker}. 
It is easily shown that this solution satisfies Eq.\ (\ref{eq10}) with $a(E)=m^2(1+\nu)$ and $b(E)=0$. 
The periodicity of the solution together with the requirement of continuity of the solution everywhere 
between the boundaries (inside the domain $0<x<\ell$) leads to the following constraint on the constants $\nu$ and $m$: $m= 2 {\bf K}(\nu)/\ell$, 
where ${\bf K}(\nu)$ represents the complete elliptic integral of the first kind with $0 \leq \nu \leq 1$ 
(as prescribed by the periodicity of the solution). 
This solution is approximately constant away from the boundaries and in the limit of large separation, 
$\ell \rightarrow \infty$, it returns the BCS solution. 
A second independent self-consistent real solution compatible with the boundary conditions is 
\begin{equation}
\Delta=m \frac{\mathsf{cn}(mx, \nu)}{\mathsf{sn}(mx, \nu)}.
\label{normaltype}
\end{equation}
This solution satisfies Eq.~(\ref{eq10}) with $a(E)=m^2(\nu-2)$ and $b(E)=0$.
A procedure analogous to the BCS solution allows to reproduce, in the limit of $\nu\rightarrow 1$ and $\ell \rightarrow \infty$, the normal state. We label the solutions Eqs.\ (\ref{bcstype}) and (\ref{normaltype}) as BCS- and normal-type, respectively. While the method adopted guarantees the solutions to be self-consistent, we have solved the system numerically and directly verified this property. Once the type of solution is fixed, the value of the parameter $\nu$ depends on the choice of the coupling constant $g$: decreasing $g$ leads to a decrease in $\nu$.

\begin{figure}
\centering
\includegraphics[width=20pc]{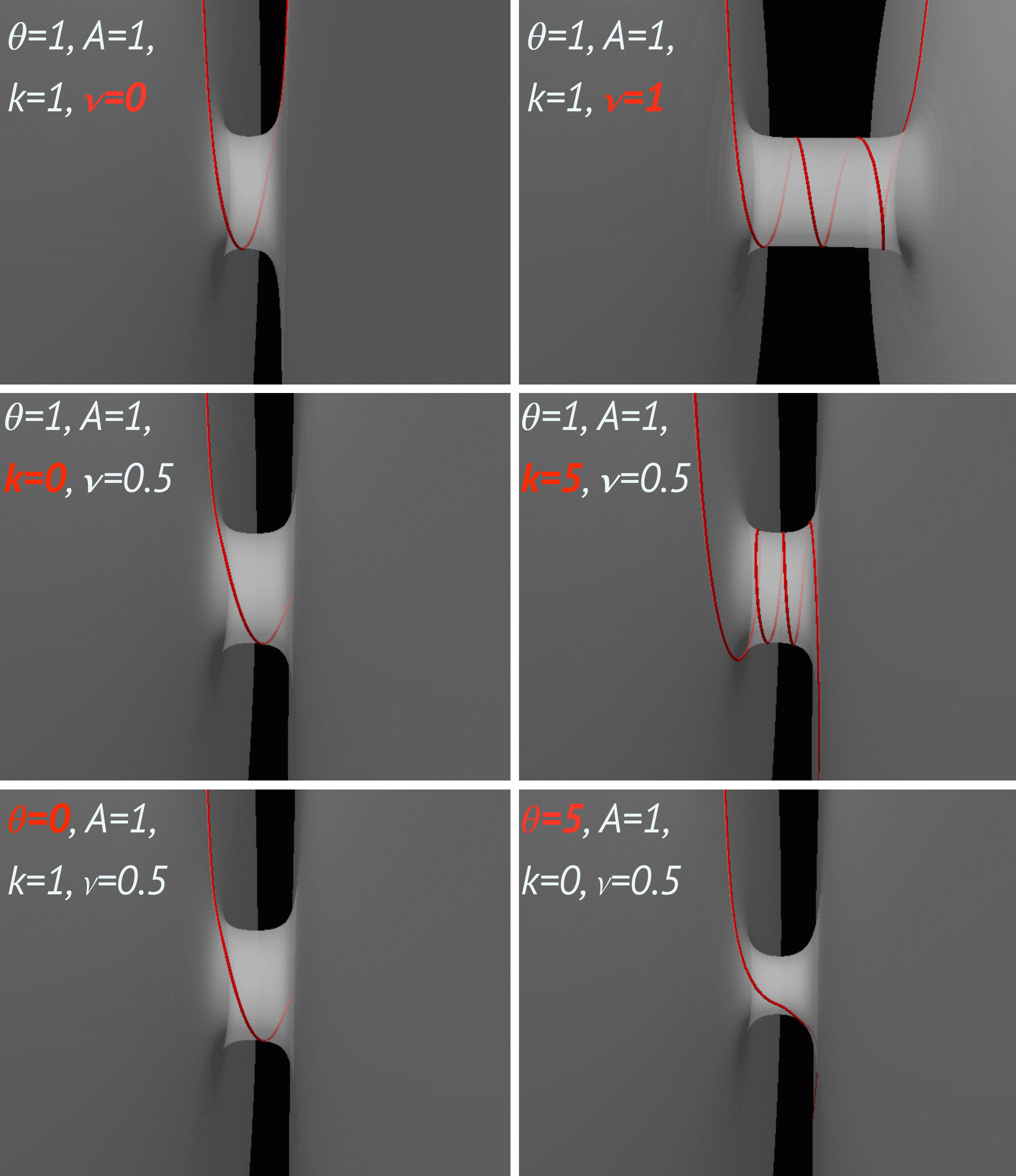}
\caption{Typical profiles of the complex solution for the condensate illustrating how the solution changes when the parameters are varied. The solid (red) line represents the phase winding.} 
\label{condensato_complesso}
\end{figure} 
For complex $\Delta$ the corresponding nonlinear Schr\"odinger equation acquires an additional term proportional to the first derivative of the order parameter [$b(E)\neq 0$]. 
Extending the previous calculation to this case is nontrivial. Details will be given elsewhere; 
here we present a general solution compatible with the boundary conditions (i.e., divergent at the boundaries), 
as the reader can directly check:
$$
\Delta = A e^{-\left[\zeta\left(\alpha{\bf K}(\nu)+i \theta/2\right)-i k\right]A x}\frac{\sigma\left(Ax+\alpha{\bf K}(\nu)+i \theta/2\right)}{\sigma\left(Ax\right)\sigma\left(\alpha{\bf K}(\nu)+i \theta/2\right)}
$$
where $\alpha=0,1$ labels the solution type ($0$ for BCS- and $1$ for normal-type). 
The quantities $\theta,~A,~\nu,~k$ are constant parameters. 
Figure \ref{condensato_complesso} shows how the solution changes under variations of these constants (animations are available in Supplemental Material \cite{Suppl}). 
The functions $\zeta$ and $\sigma$ are the Weierstrass elliptic functions with two periods $2\bm{K}(\nu)$ and $2i\bm{K}(1-\nu)$ \cite{Abramowitz,Whittaker}. 
The limit $\alpha\to0$, $\theta\to 2{\bf K}(1-\nu)$, and $k\to0$ returns the real BCS solution, while the limit $\alpha\to1$, $\theta\to 0$ and $k\to0$ returns the real normal condensate. 
The divergence at the boundaries is controlled by $\sigma(Ax)$ in the denominator, and the boundary conditions require the parameter $A$ to be quantized according to the relation $\sigma(A \ell)=0$, which forces the solution to diverge at the boundary $x=\ell$ and $A={2 {\bf K}(\nu)}/{\ell}$. 
At the other end point $x=0$, the boundary conditions are trivially satisfied $\sigma(0)=0$. 

\section{Vacuum energy} 
Using the results of the previous section, we can now calculate the vacuum energy. 
A specific feature of this setup is that the spectrum depends on which family of solutions for the condensate is selected. 
To illustrate our point, it is sufficient to  
consider the case of real condensate and select the BCS-type solution $\alpha=0$. 
We have checked that the conclusions can be extended to the other types of solutions and are generic, 
and a comprehensive analysis will appear elsewhere.
It is interesting that some limiting cases (e.g., $\nu\to0$ and real condensate) can be treated semianalytically. 
Here, in order to arrive at more general results, we will proceed numerically. 
In the present setup, the vacuum energy consists of a contribution coming from the condensate plus the proper Casimir term. 
These two, as we will shortly see, will give competing contribution to the Casimir force. In the limit of vanishing coupling (free fermions), only the latter survives. 

Since we are limiting our considerations to the leading large-$N$ approximation, 
the condensate contribution can be directly obtained by evaluating $\int_0^\ell {\Delta^2}/{(2g)}$. For $\alpha=0$, 
this can be easily calculated using 
$\int_0^\ell \Delta^2\; dx = {8\eta(\nu)}/{\ell}$, 
where $\eta(\nu)={\bm K}(\nu)[{\bm E}(\nu)-{\bm K}(\nu)]$, where ${\bm E}(\nu)$ is the complete elliptic integral of the second kind \cite{Abramowitz,Whittaker}. 
Beyond the mathematical detail, what is more important to notice here is that $\eta(\nu)\to 0$ for $\nu\to0$ and $\eta(\nu)> 0$ for any $\nu>0$. 
This term gives a {repulsive} contribution to the force. 
Also, one should notice that the scaling of the condensation energy as $\ell^{-1}$ appears nontrivially as the equation for the condensate is nonlinear. 
The same scaling occurs for the fermion contribution that we compute below. In the latter, however, the $1/\ell$ behavior is readily explained as a result of the large-$N$ approximation.

The fermion contribution comes from the summation over the energy eigenvalues. For $\alpha=0$, the spectrum can only be computed explicitly only for special values of $\nu$. 
Here, we have proceeded numerically and used the following ansatz for the spectrum 
$\mathsf{E}_n =({\pi}/{\ell})\sqrt{\left(n+{1}/{2}\right)^2 + \omega^2(\nu,n)}$. 
This form of the spectrum has been guessed by computing numerically the spectrum and by requiring that the energy eigenvalues reproduce the $\nu=0$ result. Numerical checks to confirm the validity of the ansatz were also carried out. 
The function $\omega(\nu,n)$ should satisfy the following properties: $\omega(\nu,n)\to0$ for $\nu=0$, as it follows from the explicit knowledge of the spectrum in this limit; 
$\omega(\nu,n)\sim O(n^z)$ with $z\leq0$ for large $n$ to agree with a linear dispersion relation. Numerical fitting indicates that $\omega(\nu,n) \approx \omega(\nu) + \cdots$, where $n$ dependence is found to be negligible. 

\begin{figure}
\center{\includegraphics[angle=0,width=\columnwidth]{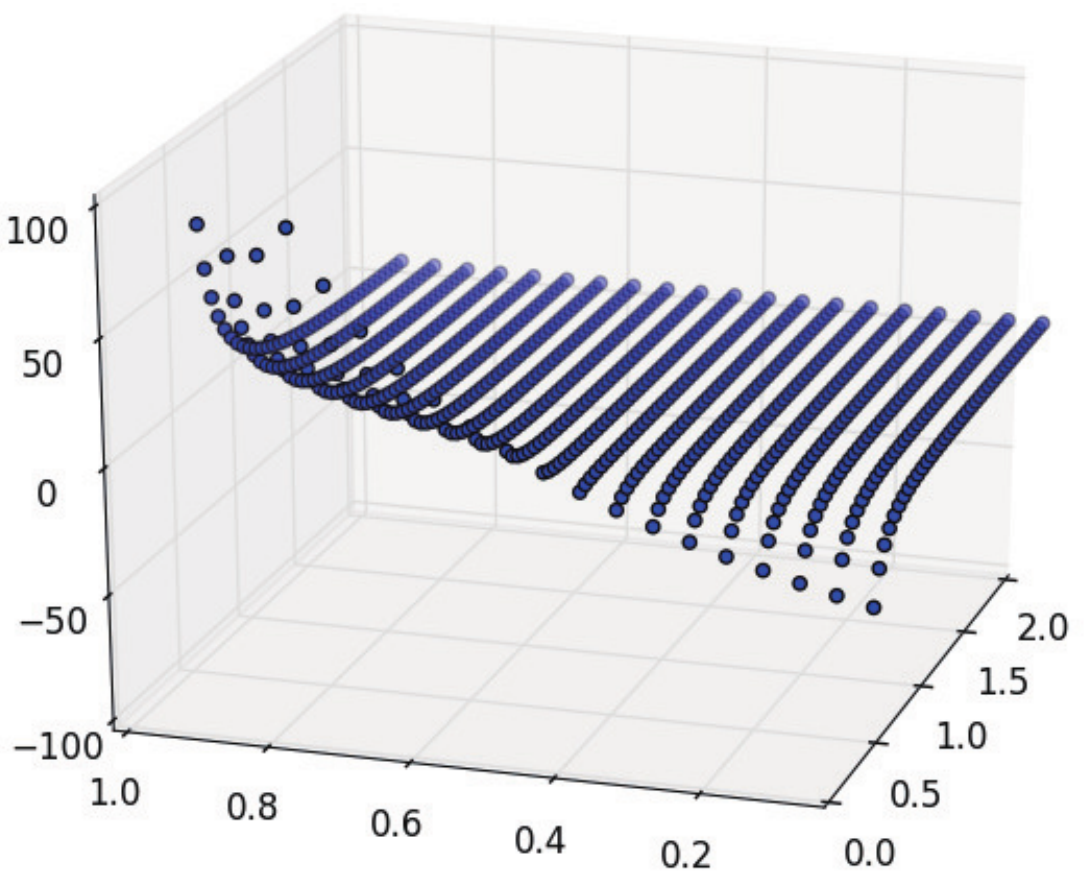}}
\put(-10,30){\Large{$\ell$}}
\put(-140,5){\Large{$\nu$}}
\put(-100,60){\textit{attractive}}
\put(-190,100){\textit{repulsive}}
\put(-250,80){\rotatebox{90}{\large{Pressure}}}
\caption{Casimir pressure (in units of $g$) 
as a function of the parameter $\nu$ and of the separation $\ell$.} 
\label{pressione}
\end{figure}
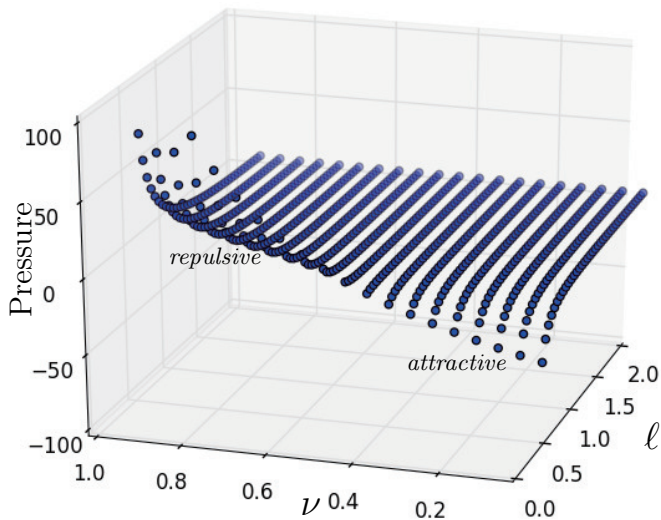 
Results of the numerical calculations are illustrated in Fig.~\ref{pressione}, where the Casimir pressure is shown  
as a function of both of the parameter $\nu$ and of the separation $\ell$. 
The sign of the force changes as a function of the parameter $\nu$ with the flip between the attractive and repulsive regimes occurring at $\nu\approx 0.4$. Reducing (increasing) the coupling $g$, reduces (increases)  $\nu$ leading to an attractive (repulsive) behavior.  

\section{Conclusions} 
In this work, we have considered a chain of fermions of length $\ell$, 
and allowed the fermions to interact, while being forced to obey boundary conditions at the edges. 
Our goal was to explore the possibility to induce transitions between attractive and repulsive regimes in the Casimir force, keeping the boundary conditions fixed. 
As a concrete example, we have used the chiral Gross-Neveu model, 
for which one expects distinct contributions to the vacuum energy coming from the condensation as well as from the original fermionic degrees of freedom. 
We have approached the problem by using the large-$N$ approximation and its reformulation in terms of a nonlinear  Schr\"odinger equation. 
Besides having shown that the solutions found in Refs.\ \cite{Basar1,Basar2,Takahashi} cease to be admissible when 
boundary conditions are enforced, 
we have clarified the near-boundary behavior of the condensate compatible with the boundary conditions and found admissible general solutions for both real and complex condensates. 
Explicit knowledge of the condensate was then used to work out the vacuum energy. 
We have found that the vacuum energy critically depends on the elliptic modulus parameter $\nu$ that, 
in turn, depends on the coupling constant $g$. 
Even for real condensates, we have observed a nontrivial behavior displaying a switch from an attractive to a repulsive regime occurring at a critical value $\nu\approx 0.4$. 
This change stems from the competition between an attractive force from the usual Casimir effect of fermions and a repulsive force from the condensate. 

It is fair to remark that, while far away from the boundaries the vacuum energy can be tuned to a negligibly small value, near the boundary the transition is not smooth, resembling a first order phase transition. This issue should be further investigated by using a confining potential rather than sharp boundary conditions. In this case, it is likely that the condensate near the boundary is regular and the transition between the attractive and repulsive regimes can possibly be smoothened out.

Here, we have considered one example of a quantum field theory on a bounded domain. A question to ask is whether the results we find here are generic to other field theory models.  It should be interesting to translate the results of our Letter to the ${\mathbb C}P^N$ model of Refs.~\cite{Milekhin:2012ca,Bolognesi:2016zjp} with which our setup shares some similarities, or looking at the example discussed in Ref.~\cite{Chernodub}.

Intuition suggests that there should be more general self-consistent solutions on the interval \cite{Takahashi:2012aw,Takahashi} or on the ring (in the presence of magnetic fields) \cite{Yoshii:2014fwa}. 
It seems worth pursuing further these analyses to clarify the role of more general boundary conditions, and how these alter the solutions for the condensate,  
the vacuum energy, and any eventual sign flip in the Casimir force.

Another amusing direction may be to investigate what happens once the model is confined within a spherical enclosure, as this may suggest possible ways of stabilization of the original Casimir mouse trap model.

Finally, perhaps the most interesting ramification of this work is to look at pistonlike configurations. In this case, it is intuitively clear that keeping the two outer boundaries fixed, an oscillatory behavior can be induced in the central defect (see Ref.~\cite{Capasso1} for a similar idea on Casimir oscillators). It is intriguing to think about the possibility of verifying this by building a cold-atomic piston or using other kinds of lower-dimensional systems. 

\section*{Acknowledgments}
The support of the Ministry of Education,
Culture, Sports, Science (MEXT)-Supported Program for the Strategic Research Foundation at Private Universities ``Topological Science" (Grant No.\ S1511006) is gratefully acknowledged. 
The work of M.~N.~is supported in part by the Japan Society for the Promotion of Science
(JSPS) Grant-in-Aid for Scientific Research (KAKENHI Grant No.~25400268).
The work of M.~N.~is also supported in part by a Grant-in-Aid for
Scientific Research on Innovative Areas ``Topological Materials
Science'' (KAKENHI Grant No.~15H05855) and ``Nuclear Matter in Neutron
Stars Investigated by Experiments and Astronomical Observations''
(KAKENHI Grant No.~15H00841) from the MEXT of Japan.


\begin{thebibliography}{99}

\bibitem{Casimir1}
H.\ B.\ G.\ Casimir, Proc.\ K.\ Ned.\ Akad.\ Wet., Ser.\ B {\bf 51}, 793 (1948).

\bibitem{Sparnay}
M.J.\ Sparnay, Physics {\bf 24}, 751 (1958).

\bibitem{Casimir2}
H.\ B.\ G.\ Casimir, Pysica (Amsterdam) {\bf XIX}, 846 (1953).

\bibitem{Boyer} 
T.H. Boyer, Phys.\ Rev.\ {\bf 174}, 1764 (1968).

\bibitem{MIT_bag_model}
A.\ Chodos, R.\ L.\ Jaffe, K.\ Johnson, C.\ B. Thorn, and V.\ F. Weisskopf, 
Phys.~Rev.~D \textbf{9}, 3471 (1974); A. Chodos, R. L. Jaffe, K. Johnson, and C. B. Thorn, Phys.~Rev.~D \textbf{10}, 2599 (1974); T. DeGrand, R. L. Jaffe, K. Johnson, J. Kiskis, Phys.~Rev.~D \textbf{12}, 2060 (1975); K.\ A.\ Milton, Phys.~Rev.~D \textbf{22}, 1441 (1980); Ann.\ Phys.\ (N.Y.) \textbf{127}, 49 (1980); \textit{ibid.} \textbf{150}, 432 (1983).

\bibitem{Johnson} 
K. Johnson, Acta\ Phys.\ Pol.\ B{\bf 6}, 865 (1975).

\bibitem{Fabinger:2000jd}
  M.~Fabinger and P.~Horava,
  Nucl.\ Phys.\ {\bf B580}, 243 (2000). 


\bibitem{FlachiCasimir} 
A. Flachi and T. Tanaka, Phys.\ Rev.\ D{\bf 80}, 124022 (2009).

\bibitem{Quach}
J.Q.\ Quach, Phys.\ Rev.\ Lett.\ \textbf{114}, 081104 (2015).

\bibitem{Decca}
R.S.\ Decca \textit{et al.},
Eur. Phys. J. C {\bf 51}, 963 (2007).

\bibitem{Capasso1}
H.B.\ Chan, V.A.\ Aksyuk, R.N.\ Kleiman, D.J.\ Bishop, and F.\ Capasso, Phys.\ Rev.\ Lett.\ {\bf 87} 211801 (2001).

\bibitem{dietrich}
M.\ Tr\"ondle, S.\ Kondrat, A.\ Gambassi, L.\ Harnau, and S.\ Dietrich, J.\ Chem.\ Phys.\ \textbf{133}, 074702  (2010) 

\bibitem{biocas}
B.B.\ Machta, S.L.\ Veatch, and J.P. Sethna, Phys.\ Rev.\ Lett.\ \textbf{109}, 138101 (2012).


\bibitem{Experiments}
S.~K.\ Lamoreaux, Phys.\ Rev.\ Lett.\ {\bf 78}, 5 (1997); U.\ Mohideen and A.\ Roy, Phys.\ Rev.\ Lett.\ {\bf 81}, 4549 (1998); G.\ Bressi, G.\ Carugno, R.\ Onofrio, and G.\ Ruoso, Phys.\ Rev.\ Lett.\ {\bf 88}, 041804 (2002); R.\ S.\ Decca, D.\ Lopez, E.\ Fischbach, and D.\ E.\ Krause, Phys.\ Rev.\ Lett.\ {\bf 91}, 050402 (2003).


\bibitem{Kardar:1997cu}
  M.~Kardar and R.~Golestanian,
  Rev.\ Mod.\ Phys.\  {\bf 71}, 1233 (1999).
  
\bibitem{Milton}
K.~A.~Milton, J.\ Phys.\ A {\bf 37}, R209 (2004).

\bibitem{textbook} 
M.\ Bordag, G.\ L.\ Klimchitskaya, U.\ Mohideen, and V.\ M.\ Mostepanenko, {\it Advances in the Casimir Effect}, (Oxford University Press, Oxford, 2009). 

\bibitem{Klimchitskaya}
G.\ L.\ Klimchitskaya, U.\ Mohideen, and V.\ M.\ Mostepanenko,
Rev.~Mod.~Phys. \textbf{81}, 1827 (2009).


\bibitem{Albrecht}
S.\ Reynaud, A.\ Lambrecht, Quantum Optics and Nanophotonics, (Oxford University Press, Oxford, 2013). 
  
  
  
\bibitem{Weinberg:1988cp}
  S.~Weinberg,
  Rev.\ Mod.\ Phys.\  {\bf 61}, 1 (1989).
  
  
  
\bibitem{Capasso}
H.\ B.\ Chan, V.\ A.\ Aksyuk, R.\ N.\ Kleiman, D.\ J.\ Bishop, F.\ Capasso, Science {\bf 291}, 1941 (2001); 
see F.\ Capasso {\it et al., in Casimir Physics}, edited by D.~Dalvit, P.~Milonni, D.~Roberts, F.~da Rosa (Eds.), 
(Springer-Verlag, Berlin, 2011), p.\ 249; R.\ Decca {\it et al., ibid}., p.\ 287.

\bibitem{Genet}
C.~Genet, A.~Lambrecht, and S.~Reynaud, Eur.\ Phys.\ J.\ Special\ Topics  {\bf 160}, 183 (2008).  

\bibitem{Munday}
J.\ N.\ Munday, F.\ Capasso, and V.\ A.\ Parsegian, Nature (London) {\bf 457}, 170 (2009).

\bibitem{KennethBachas}
O. Kenneth and I. Klich, Phys. Rev. Lett. {\bf 97}, 160401 (2006); C.P. Bachas, J. Phys. A {\bf 40}, 9089 (2007).

\bibitem{Schaden}
M. Schaden, Phys. Rev. Lett. {\bf 102}, 060402 (2009); see also M. Schaden, Phys. Rev. A {\bf 73}, 042102 (2006).  

\bibitem{Gies}
H. Gies, K. Langfeld, and L. Moyaerts, 
J.\ High Energy Phys.\ 06 (2003) 018.

\bibitem{Asorey}
M.~Asorey and J.~M.~Munoz-Castaneda, Nucl.~Phys. \textbf {B874} 852 (2013). 

 \bibitem{Flachi:2013bc} 
  A.~Flachi,
  Phys.\ Rev.\ Lett.\  {\bf 110}, 060401 (2013);
  Phys.\ Rev.\ D {\bf 86}, 104047 (2012).

\bibitem{Schecter}
M.~Schecter and A.~Kamenev,
Phys.\ Rev.\ Lett.\ \textbf{112}, 155301 (2014).

\bibitem{CasFer}
E.\ Elizalde, M.\ Bordag, and K.\ Kirsten, J.\ Phys.\ A: Math.\ Gen.\ {\bf 31}, 1743 (1998);
A.~Bulgac and A.~Wirzba, Phys.\ Rev.\ Lett.\ \textbf{87}, 120404 (2001);
A.\ Erdas, Phys.\ Rev.\ D \textbf{83}, 025005 (2011); 
E.\ Elizalde, S.\ D.\ Odintsov, and A.\ A.\ Saharian, Phys.\ Rev.\ D \textbf{83}, 105023 (2011);
S.\ Bellucci and A.\ A.\ Saharian, Phys.\ Rev.\ D \textbf{79}, 085019 (2009);
L.~P.~Teo, Phys.\ Rev.\ D {\bf 91}, 125030 (2015); Phys.\ Rev.\ D {\bf 91}, 085034 (2015); 
A.~Flachi and L.~P.~Teo, Phys.\ Rev.\ D {\bf 92} 025032 (2015);
P.\ Sundberg, R.\ L.\ Jaffe, Ann.~Phys.~\textbf{309}, 442 (2004);
A.~Recati, J.~N.~Fuchs, C.~S.~Pe\c{c}a, and W.~Zwerger, Phys.~Rev.~A \textbf{72}, 023616 (2005); 
J.~N.~Fuchs, A.~Recati, and W.~Zwerger, Phys.~Rev.~A \textbf{75}, 043615 (2007); 
E.~B.~Kolomeisky, J.~P.~Straley, and M.~Timmins, Phys.\ Rev.\ A \textbf{78}, 022104 (2008); 
D.\ Zhabinskaya, J.\ M.\ Kinder, and E.\ J.\ Mele, Phys.\ Rev.\ A \textbf{78}, 060103 (2008); 
D.\ Zhabinskaya and E.\ J.\ Mele, Phys.\ Rev.\ B \textbf {80}, 155405 (2009); 
M.\ Napi\'orkowski and J. Piasecki, J.\ Stat.\ Phys.\ \textbf{156}, 1136 (2014); 
P.\ Jakubczyk, M.\ Napi\'orkowski, and T.\ S\c{e}k, Eur.\ Phys.\ Lett.\ \textbf{113} 30006 (2016).


\bibitem{NJL}
Y.\ Nambu and G.\ Jona-Lasinio, Phys.~Rev.~\textbf{122}, 345 (1961); Phys.~Rev.~\textbf{124}, 246 (1961). 

\bibitem{Gross}  
D.~J.~Gross and A.~Neveu, Phys.~Rev.~D \textbf{10}, 3235 (1974). 

\bibitem{polyacetylene}
H.~Takayama, Y. R.~Lin-Liu, and K.~Maki, Phys.~Rev. B \textbf{21} (1980) 2388; 
A.~Chodos and H.~Minakata, \textit{Field Theoretical Tools for Polymer and Particle Physics}, in Lecture Notes in Physics, Vol.~508 (Springer-Verlag, Berlin, 1998).

\bibitem{supcond}
K.~Machida and H.~Nakanishi, Phys.~Rev. B \textbf{30}, 122 (1984); 
J. Bar-Sagi and C.G. Kuper, Phys.~Rev.~Lett. \textbf{28},  1556 (1972).

\bibitem{Thies:2006ti}
M.~Thies, J.\ Phys.\ A {\bf 39}, 12707 (2006).

\bibitem{Basar1} 
G.~Ba\c{c}ar and G.~V.~Dunne, Phys.~Rev~Lett.~\textbf{100}, 200404 (2008); Phys.~Rev.~D \textbf {78}, 065022 (2008).

\bibitem{Basar2} 
G.~Basar, G.~V.~Dunne, and M.~Thies, Phys.\ Rev.\ D {\bf 79}, 105012 (2009).

\bibitem{Takahashi}
D.~A.~Takahashi and M.~Nitta, Phys.~Rev.~Lett. \textbf{110}, 131601 (2013).

\bibitem{Witten}
E.\ Witten, Nucl.\ Phys.\ {\bf B145}, 110 (1978). 

\bibitem{Abramowitz}
\textit{Handbook of Mathematical Functions with Formulas, Graphs, and Mathematical Tables}, 
edited by M.~Abramowitz and I.~A.~Stegun, (Dover, New York, 1972).

\bibitem{Whittaker}
E.~T.~Whittaker and G.~N.~Watson, \textit{A Course in Modern Analysis}, (Cambridge University Press, Cambrigde, England, 1990).

\bibitem{Milekhin:2012ca} 
A.~Milekhin, Phys.\ Rev.\ D {\bf 86}, 105002 (2012); Phys.\ Rev.\ D {\bf 95}, 085021 (2017).

\bibitem{Bolognesi:2016zjp}
S.~Bolognesi, K.~Konishi, and K.~Ohashi, J.\ High Energy Phys.\ 10 (2016) 073. 
 
\bibitem{Chernodub}
M.N. Chernodub, V.A. Goy, and A. V. Molochkov, Phys.\ Rev.\ D {\bf 95}, 074511 (2017).

\bibitem{Takahashi:2012aw} 
D.~A.~Takahashi, S.~Tsuchiya, R.~Yoshii, and M.~Nitta, Phys.\ Lett.\ B {\bf 718}, 632 (2012).

\bibitem{Yoshii:2014fwa} 
R.~Yoshii, S.~Takada, S.~Tsuchiya, G.~Marmorini, H.~Hayakawa, and M.~Nitta, Phys.\ Rev.\ B {\bf 92}, 224512 (2015).

\bibitem{Suppl}
See ancillary files for the parameter dependence of the solution.


\end{thebibliography}
\end{document}